\documentclass[twocolumn,prb]{revtex4}
\usepackage{graphicx}
\usepackage{epsfig}
\usepackage{dcolumn}
\usepackage{bm}
\usepackage{amsmath}

\begin{document}

\title{Coherent optical control of spin-spin interaction in
doped semiconductors}

\author{C. Piermarocchi}
\author{G. F. Quinteiro}

\affiliation{Department of Physics and Astronomy, Michigan State
University, East Lansing, Michigan 48824-2320}

\date{\today}

\begin{abstract}
We provide a theory of laser-induced interaction between spins
localized by impurity centers in a semiconductor host. By solving
exactly the problem of two localized spins interacting with one
itinerant exciton, an analytical expression for the induced spin-spin
interaction is given as a function of the spin separation, laser
energy, and intensity. We apply the theory to shallow neutral donors
(Si) and deep rare-earth magnetic impurities (Yb) in III-V
semiconductors. When the photon energy approaches a resonance related
to excitons bound to the impurities, the coupling between the
localized spins increases, and may change from ferromagnetic to
anti-ferromagnetic. This light-controlled spin interaction provides a
mechanism for the quantum control of spins in semiconductors for
quantum information processing; it suggests the realization of spin
systems whose magnetic properties can be controlled by changing the
strength and the sign of the spin-spin interaction.
\end{abstract}

\maketitle

% --------------------------------------------------------------
% --------------------------------------------------------------
% --------------------------------------------------------------

\section{Introduction}
The possibility of creating semiconductor systems that can work
simultaneously as electronic, photonic and magnetic devices has
boosted recently the research on light-spin interaction in
semiconductors. These efforts extend both in the direction of an
optical control of macroscopic magnetic properties and toward the
quantum control of single spins.~\cite{awschalom02} Examples belonging
to the first class include the investigation on light-induced
paramagnetic to ferromagnetic transitions in magnetic semiconductors,
using coherent~\cite{fernandez03} or incoherent~\cite{boukari02}
processes, and the attempts to control the macroscopic spin
polarization of carriers using polarized light.~\cite{ohno98} More
recently, schemes for optical control at the level of few spins have
been proposed. Bao et al.~\cite{bao03} have demonstrated
quantum spin entanglement of few donors and magnetic impurities. The
strong potentialities of this optical quantum control of spins for
quantum information processing using quantum
dots~\cite{piermarocchi02,pazy03} and impurities~\cite{stoneham03}
have been emphasized, and recent advances in the optical control at
the nanoscale in semiconductor nanostructures~\cite{li03,bianucci04}
are particularly encouraging in this perspective.

It was pointed out in Ref.~\onlinecite{piermarocchi02} that itinerant
excitons, i.e. optical excitations free to move in the host material
that embeds the localized spins, can induce an effective spin-spin
interaction between localized spins. This mechanism has been dubbed
{\it Optical RKKY} (ORKKY), in analogy to the mechanism in theory of
magnetism,~\cite{rudeman54} where electrons are involved. In the
coherent optical case virtual excitons are created, and the ORKKY
coupling is obtained from a second order perturbation theory in the
exchange coupling between the itinerant exciton and the localized
spin.  The ORKKY result predicts that the coupling between the
localized spins is always ferromagnetic, independently of the sign of
the coupling with the excitons. In this paper we show that higher
order terms in the exciton-impurity coupling can modify the strength
and sign of the interaction, and affect its dependence as a function
of the spin separation.  The calculation of the spin-spin interaction
can be reduced to a spin dependent scattering problem that can be
solved including exactly all the multiple scattering terms between the
two localized spins.  We follow here an approach similar to that used
to calculate multiple scattering effects of $\pi^{\pm}$ mesons by
deuterons.~\cite{drell55} The higher order terms in the
exciton-impurity coupling describe bound states which affect strongly
the optically induced spin-spin interaction.  In particular, a
controlled anti-ferromagnetic (AF) coupling can be realized when the
laser energy is tuned in the bonding-antibonding gap for the exciton
localized by two impurities. This laser controlled switching of sign
of the spin-spin interaction opens to new directions in the
investigation of competing interactions in spin systems.
\begin{figure}[b]
\vspace{-0.8 cm}
\centerline{\includegraphics[scale=0.37]{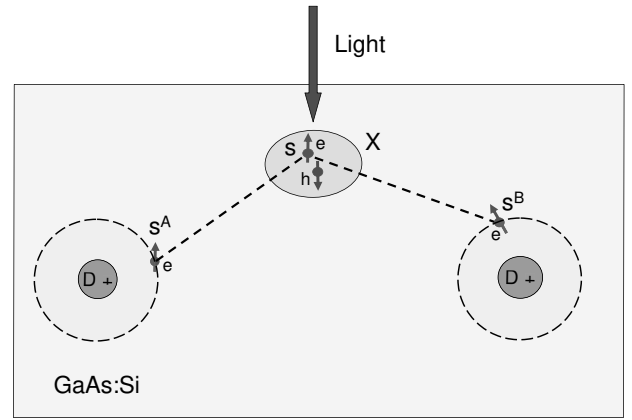}}
\vspace{-1.cm}
\caption{Scheme of the light induced spin-spin interaction in the case
of two shallow donors. \label{fig0}}
\end{figure}

The extension of the RKKY to insulators was originally proposed by
Bloembergen and Rowland.~\cite{bloembergen55} They predicted the
exponential decay of the spin-spin interaction with a characteristic
length $\kappa=\hbar/\sqrt{2 m E_g}$ depending on the energy gap $E_g$
and mass of the virtual electron hole pairs across the gap. In the
optically induced RKKY, the energy gap is effectively reduced by the
laser field which increases the effective length to
$\kappa=\hbar/\sqrt{2 m (E_g-\hbar\omega_P)}$, with $\hbar \omega_P$
being the energy of the laser. Also, the density of electrons in the
occupied bands in the insulator is replaced in the optical case by the
density of photons in the field. The innovative strength of the
optically induced case resides in the control potentialities since
both the intensity and the frequency of the laser can be controlled in
an experiment. There are intrinsic difficulties to go beyond second
order perturbation theory in the case of metals.~\cite{vertogen67}
These difficulties are not present in the optical coherent case since
there is no Fermi sea of electrons. The presence of a Fermi sea
simultaneously with the laser would produce light-induced Kondo
effects,~\cite{perakis00} which we do not consider here.

The paper is organized as follows: In section~\ref{effham} we derive
the expression for the effective Hamiltonian of two localized spins in
the presence of a light field, and we relate it to the spin-dependent
T matrix operator of a two-center scattering problem. We study first
in Sec.~\ref{one} the scattering of one exciton with one center. By
generalizing a result from scattering theory~\cite{mott65} to the spin
dependent case, we show in section~\ref{sect-twospin} how the T matrix
operator for the exciton scattering on two centers can be expressed in
terms of the T operator for the one center scattering. We also study
in this section the effects of the polarization of the light and we
show that a circularly polarized field will induce an additional term
representing a magnetic field. The theory is applied in
section~\ref{app} to two systems: shallow donors, and deep rare earth
magnetic impurities. We discuss implications for quantum computing
implementations and for optical control of macroscopic magnetic
properties in Sec.\ref{qc}.

% --------------------------------------------------------------
% --------------------------------------------------------------
% --------------------------------------------------------------

\section{Effective hamiltonian for localized spins coupled by the light}
\label{effham}
A schematic view of the effect in the case of two shallow donors is
given in Fig.~\ref{fig0}.  We are not interested in calculating the
optical properties of the whole system, but we want to consider the
effect of a coherent field on the dynamics of the two non-interacting
localized spins $s^A$ and $s^B$. The light creates virtual/real
excitons in the semiconductor host and affects the localized spin
states. Due to their coupling to the light, the localized spins can therefore
interact with each other. We want to study the behavior of the two
localized spins in the coherent optical regime. This implies that the
laser is always off resonance with respect to the free exciton band to
avoid strong energy absorption. We therefore consider only single
exciton processes in the presence of a monochromatic laser field.  The
system of two localized spins coupled to one itinerant exciton is
described by the Hamiltonian
\begin{equation}
{H}_{XS}=H_0+H_1~,
\label{h1}
\end{equation}
where, $H_0$ describes a free exciton of mass $M$ with dispersion
$\epsilon_k= \epsilon_0+\hbar^2 k^2/2M $. The term $H_1$ describing the spin
dynamics can be written in the form
%\begin{widetext}
\begin{eqnarray}
H_1&=&\frac{1}{V}\sum_{k k^{\prime}\alpha \alpha^{\prime}\beta }
J_{k,k^{\prime}}(s^A \cdot s_{\alpha^{\prime}
\alpha} \nonumber\\
&+&{e^{-i({k^{\prime }}-k)R}}~s^B\cdot s_{\alpha^{\prime}
\alpha})b^{\dagger}_{k^{\prime} \alpha^{\prime} \beta} b_{k \alpha
\beta}
\label{ham_H1}
\end{eqnarray}
%\end{widetext}
where the two localized spins $\frac{1}{2}$ are described by $s^{A}$
and $ s^{B}$. $V$ is the volume, and $s$ is the electronic spin of the
itinerant exciton. $b^{\dagger}_{k \alpha \beta}$ creates an exciton
with center of mass momentum $k$, electron spin $\alpha$, and hole
spin $\beta$. $R$ is the separation between the two impurities.
$J_{k,k^{\prime}}$ is the exciton-spin exchange interaction. The
strength and the sign of this term depend strongly on the nature of
the localized spin. The sign, for instance, is determined by the
competition between the ferromagnetic potential exchange and the
antiferromagnetic kinetic exchange which is due to the hybridization
of the itinerant exciton state with the localized
state.~\cite{anderson63} We will keep for the moment a general
approach independent of the nature of the $J_{k,k^\prime}$~, and we
will discuss two specific examples in section \ref{app}.  A
spin-independent term corresponding to a direct Coulomb interaction
between the exciton and the impurity is also present. This term is
small for shallow impurities, where kinetic exchange effects dominate,
but becomes important for deep impurities. We will include this term
in the case of rare earth impurities discussed in Sec.~\ref{Yb}, and
we disregard it in the general discussion since it only introduces
spin-independent energy shifts.  We assume that the $1s$ excitons
dominate the light induced effect, as discussed in
Ref.~\onlinecite{piermarocchi02}. Moreover, we focus on systems where
the localized states interact only with the electron in the exciton:
the full Hamiltonian in Eq.~(\ref{h1}) is diagonal in the hole spin
index $\beta$. This is a good approximation for electrons in neutral
donors, since it is equivalent to neglecting the electron-hole exchange
interaction which in most semiconductors is much smaller than the
electron-electron exchange.  Concerning the second example we will
consider, i.e. the case of the Yb$^{3+}$ ions in III-V, it is known
that these ions act as strong isoelectronic traps for electrons and
the $s$-$f$ exchange in the conduction band dominates.

The interaction of the excitons with an external time
dependent optical field provides the mechanism for the control of the
two localized spins and is described by the Hamiltonian
\begin{equation}
H_{XL}=\sqrt{V}\sum_{\sigma} \Omega_{\sigma} e^{i \omega_L t}
 \phi_{1s} b_{k=0,\alpha+\beta=\sigma}+ hc~,
\label{hlaser}
\end{equation}
where $\Omega_{\sigma}$ is the Rabi energy of the interband optical
transition and $\hbar\omega_L$ is the energy of the laser, $\sigma$ is
the polarization of the light. We have used the rotating wave
approximation in Eq.~(\ref{hlaser}). $\phi_{1s} $ is the envelope
function of the electron hole pair taken at $\rho=r_e-r_h=0$. In the
case of a cw laser field, the time dependence can be eliminated by
moving to the rotating frame with frequency $\omega_L$, thus replacing
$\epsilon _k$ by $\epsilon _k-\hbar\omega_L$ in $H_0$.

We are deriving an effective Hamiltonian for the two localized spins
in the presence of the laser field.  For a fixed value of the
polarization $\sigma$ this Hamiltonian is four dimensional,
corresponding to the degenerate ground state described by
$|\lambda\rangle=\{|\uparrow\uparrow\rangle,
|\uparrow\downarrow\rangle, |\downarrow\uparrow\rangle,
|\downarrow\downarrow\rangle\}$ for the two localized spins and can be
written to the second order perturbation theory in $H_{XL}$ as
\begin{equation}
H_{\lambda\lambda^\prime}^{eff}=\langle\lambda|H_{XL}
\frac{1}{\varepsilon_\lambda^0-PH_{XS}P} H_{XL}|\lambda^\prime\rangle
\end{equation}
where $P$ is the projector operator onto the subspace of one exciton
plus two spins and $\varepsilon_\lambda^0$ is the ground state energy
of two spins with no exciton.  Operating $H_{XL}$ onto $\mid
\lambda\rangle$ generates states with one exciton at $k=0$ with
polarization $\sigma$, states written as \{ $\mid\lambda 0
\sigma\rangle$ \}. The ground state energy can be chosen to be
$\varepsilon_\lambda^0=0$, and the expression above simplifies to
%\begin{widetext}
\begin{equation}
H_{\lambda\lambda^\prime}^{eff}=V
\sum_{\sigma}\mid\Omega_\sigma\mid^2\mid\phi_{1s}\mid^2 G_{\lambda 0
\sigma,\lambda^{\prime} 0\sigma}(\omega_L) \label{h2}
\end{equation}
%\end{widetext}
where 
\begin{equation}
G_{\lambda 0 \sigma,\lambda^{\prime} 0\sigma}(\omega_L) =\langle
\lambda 0 \sigma
|\frac{1}{[G^0(\omega_L)]^{-1}-H_{1}}|\lambda^{\prime} 0 \sigma
\rangle
\end{equation}
is the Green's function operator for the system composed by the
exciton and two spins, and
\begin{equation}
G^0(\omega_L)=\frac{\delta_{kk^\prime}}{\hbar \omega_L-(\epsilon_0
+\frac{\hbar^2k^2}{2M})+ i\eta}~.
\end{equation}
We remark that we work in the off-resonance regime for which
$\omega_L<\epsilon_0$, thus making the real part of $G^0$ always
negative. Since we are dealing with only single-exciton processes, the
Lippman-Schwinger equation for $G$ can be rewritten in terms of an
equation for the $T$-matrix defined by the relation $G=G_0+G_0 T
G_0$. We solve the problem in two steps: $(i)$ the $T^A$ and $T^B$
operators representing the scattering of the exciton with only one
impurity (identified by the index $A$ or $B$) is solved. $(ii)$ The
T-matrix for the exciton interacting with two impurities is explicitly
rewritten in terms of $T^A$ and $T^B$ using \cite{mott65}
\begin{equation}
T=\frac{1}{1-T^A G^0 T^B G^0} T^A[1 +G^0 T^B]+(A \rightleftharpoons B)
\label{h3}
\end{equation}
where $(A \rightleftharpoons B)$ stands for repeating the previous
term with interchange of superscripts $A$ and $B$.  Eq.~(\ref{h3})
takes into account exactly all the multiple scattering processes
between the exciton and the two localized spins.  We will focus in the
next section on the interaction of the exciton with a single localized
spin. The role of multiple scattering effects in the two spins case
are addressed in section \ref{sect-twospin}.

\section{Exciton-single impurity scattering}
\label{one}
This section focuses on the solution of the T-matrix equation for one
scattering center (named $A$). Due to the short range nature of the
exchange interaction, the exchange integral $J_{k,k^\prime}$ in
Eq.~(\ref{ham_H1}) is often reduced to a constant, corresponding to a
delta-like interaction in space. Here we consider a more realistic
form of the interaction using the separable potential
approximation~\cite{yamaguchi54,schick61} where $J_{k,k^{\prime}}=J
v_k v_{k^{\prime}}$, with $v_k$ being a dimensionless form factor that
depends only on $k=|\vec{k}|$. $v_k$ describes the effect of the
finite size of the non-local exchange interaction. The separable form
of $J_{k,k^{\prime}}$ will allow us to obtain analytical expressions
for the T-matrix, and provide a flexible theoretical framework with
parameters that can be taken for the experiments. On the other hand,
this potential can support at most one s-like bound state.  The
integral equation for the T-matrix, $T=H_1+H_1 G_0 T $, can then be
written explicitly as
%\begin{widetext}
\begin{eqnarray}
T^A_{kk^{\prime}\alpha\alpha^{\prime}}=\frac{J}{V} v_kv_{k^{\prime}}s^A \cdot
s_{\alpha\alpha^{\prime}}+ \nonumber \\ +
\frac{J}{V}\sum_{k^{\prime\prime}\alpha^{\prime\prime}}v_kv_{k^{\prime\prime}}s^A\cdot
s_{\alpha\alpha^{\prime\prime}} G^0_{k^{\prime\prime}}
T^A_{k^{\prime\prime}k^{\prime}\alpha^{\prime\prime}\alpha^{\prime}}~.
 \label{h5}
\end{eqnarray}
%\end{widetext}
We can write the T-matrix as a sum of a scalar and vector part 
\begin{equation}
T^A_{kk^{\prime}\alpha\alpha^\prime}=\frac{v_k v_{k^{\prime}}}{V}[T_0
\delta_{\alpha\alpha^{\prime}}+T_1 s^A \cdot
s_{\alpha\alpha^{\prime}}]~,
%\doteq v_k
%v_{k^{\prime}}\hat{\Upsilon}^A_{\alpha,\alpha^\prime}
\label{Tsol}
\end{equation}
and, using the identity 
\begin{equation}
(s^A \cdot
s)^2=\frac{3}{16}-\frac{s^A \cdot
s}{2}~,
\end{equation}
we rewrite Eq.~(\ref{h5}) in the form of two coupled equations
\begin{subequations}
\label{coupledT}
\begin{eqnarray}
T_1&=&J+JF_0 (T_0-T_1/2) \\
T_0&=&\frac{3}{16}J F_0 T_1
\end{eqnarray}
\end{subequations}
where
\begin{equation}
F_0(\omega_L)=\frac{1}{V}\sum_{k^{\prime\prime}}
v_{k^{\prime\prime}}^2G^0_{k^{\prime\prime}}~. \label{f0}
\end{equation}
The reduction of the integral equation to two  algebraic equations
is a consequence of the form of the interaction. The two coupled
equations in Eq.~(\ref{coupledT}) are solved and give
\begin{subequations}
\label{const}
\begin{eqnarray}
T_0&=&\frac{3J}{16}\frac{JF_0}{1+\frac{J
F_0}{2}-\frac{3(J F_0)^2}{16}} \\
T_1&=&\frac{J}{1+\frac{J F_0}{2}-\frac{3(J F_0)^2}{16}}~.
\end{eqnarray}
\end{subequations}
%e^{ik^{\prime\prime}\cdotR}
This analytical solution allows us to investigate the strong coupling
regime in which the quantity $JF_0$ is not small. The most interesting
feature of the strong coupling regime is the formation of bound
states of the exciton with the impurities, identified by the
poles in the T-matrix. Varying the frequency of the laser $\omega_L$,
which will modify the $F_0$, we can scan the spectrum to obtain the
energy of those bound states. We remark that under the condition
$\omega_L<\epsilon_0$, and assuming reasonably that that the potential
$v(k)$ is an analytic function of $k$, no singularities or branch cuts
exist for the function $F_0$. Therefore, the only source of poles is
given by the zeros of the function $1+\frac{JF_0}{2}-\frac{3
(JF_0)^2}{16}$, appearing in $T_1$ and $T_0$. Considering separately
the singlet and triplet channels we find
\begin{subequations}
\label{ST}
\begin{eqnarray}
T^{S}&=&-\frac{3/4 J}{1+ 3/4 J F_0} \label{ts} \\
T^{T}&=&\frac{J/4}{1- 1/4 J F_0}~.
\end{eqnarray}
\end{subequations}
In the singlet and triplet channels only one of the two poles in
Eq.~(\ref{const}) is present. Due to the fact that $F_0(\omega_L)$ is
negative for all allowed values of $\omega_L$, we also remark that, as
expected, the exciton binds in a singlet spin state if $J>0$
(antiferromagnetic coupling), while the bound state results to be a
triplet if the exciton-electron exchange is ferromagnetic
(i.e. $J<0$).

% --------------------------------------------------------------
% --------------------------------------------------------------
% --------------------------------------------------------------

\section{Exciton-two impurities scattering}
\label{sect-twospin}
Starting from the results obtained in the previous sections, we
construct in this section the solution for the exciton-two impurities
T-matrix and the corresponding $H_{eff}$ for the localized
spins. Eq.~(\ref{h3}) can be expanded in terms of $T$ operators as
\begin{equation}
T=T^A+T^A G^0 T^B+T^A G^0 T^B G^0 T^A+...
\label{seriesT}
\end{equation}
The matrix for $T^B$ can be obtained from a simple phase shift: if
$T^A_{kk^{\prime}}$ is the T-matrix for a scattering center with
potential $V(r)$ then $e^{-i(k^{\prime}-k)\cdot R} T^A_{kk^{\prime}}$
is the corresponding one for a potential $V(r-R)$,
i.e. $T^B_{kk^{\prime}}$.~\cite{mott65}  We can take the matrix elements of
Eq.~(\ref{seriesT}) in the $k$ representation. To illustrate  how
this series can be summed let us consider as an example the third term
in Eq.~(\ref{seriesT})
\begin{widetext}
\begin{equation}
\langle k \mid T^A G^0 T^B G^0 T^A \mid k^\prime \rangle=\frac{1}{V^3}\sum_{k^{\prime\prime},k^{\prime\prime\prime}}
v_{k}v_{k^{\prime\prime}} \Upsilon^A G^0_{k^{\prime\prime}}
e^{-i(k^{\prime\prime\prime}-k^{\prime\prime})\cdot
R}v_{k^{\prime\prime}}v_{k^{\prime\prime\prime}}  \Upsilon^B
G^0_{k^{\prime\prime\prime}}v_{k^{\prime\prime\prime}}v_{k^\prime}
\Upsilon^A~
\end{equation}
\end{widetext}
where we have defined
\begin{equation}
\Upsilon^{A(B)}=T_0+T_1 s^{A(B)}\cdot s~.
\end{equation}
Reordering factors and defining the function 
\begin{equation}
F_R(\omega_L)=\frac{1}{V}\sum_{k} e^{i k \cdot R} v^2_k G^0_{k}~,
\label{fr}
\end{equation}  this term takes the form,
\begin{equation}
\frac{v_k v_{k^\prime}}{V}F_R^2(\omega_L) \Upsilon^A
\Upsilon^B \Upsilon^A~. \label{secondtermT}
\end{equation}
Following the same procedure, the full series can be
summed to,
\begin{equation}
T_{k,k^{\prime}}=\frac{v_k v_{k^\prime}/V}{1-F_R^2 \Upsilon^A
\Upsilon^B} \Upsilon^A [1+F_R\Upsilon^B]+(A
\rightleftharpoons B)~. \label{finalT}
\end{equation}
The $T$-matrix is now expressed as an operator in a 8-dimensional
space generated by three spins $\frac{1}{2}$: one electron in the exciton and
two localized electron states.  By direct inversion and products of
8 by 8 matrices, Eq.~(\ref{finalT}) can be rewritten in terms of a
combination of spin products (see Appendix \ref{appendix}), and using
$G=G_0+G_0 T G_0$ we obtain the spin dependent effective Hamiltonian
\begin{equation}
\label{Heff}
H_{eff}=B_{eff}\cdot(s^A+s^B) +J_{eff} s^A\cdot s^B~.
\end{equation}
$B_{eff}$ represents an effective magnetic field acting on both 
spins and $J_{eff}$ is an effective isotropic Heisenberg exchange. The
effective magnetic field and exchange constant can be written as
\begin{widetext}
\begin{equation}
B_{eff}=
\frac{|\Omega_{\sigma+}|^2-|\Omega_{\sigma-}|^2}{\delta^2}\frac{|\phi_{1s}|^2
v^2_0 J(1-J F^{-}_R)}{(1-JF^+_R)[1-J F^+_R(3 J F^-_R
-2)]}\frac{\hat{z}}{2}
\label{beff}
\end{equation}   
and
\begin{equation}
J_{eff}=
\frac{|\Omega_{\sigma+}|^2+|\Omega_{\sigma-}|^2}{\delta^2}\frac{|\phi_{1s}|^2
v^2_0 J^2 F_R/2(1-J F^-_R)}{(1-JF^+_R)[1-J F^+_R(3 J F^-_R -2)] [1-J
F^-_R(3 J F^+_R -2)]}~,
\label{jeff}
\end{equation} 
\end{widetext}   
where we have defined 
\begin{equation}
F^{\pm}_{R}(\omega_L)=\frac{1}{4 V}\sum_{k}
(1\pm e^{i k \cdot R}) v^2_k G^0_{k}~.  
\end{equation}
$\hat{z}$ identifies the direction of propagation of the light,
$\delta=\epsilon_0-\hbar \omega_L$ is the optical detuning, and
$\Omega_{\sigma\pm}$ correspond to the contributions to the Rabi
energy from the two circularly polarized components of the light.
From Eq.~(\ref{finalT}) a spin-independent term is also derived which
is not shown in Eq.~(\ref{Heff}) since it is irrelevant for our
purposes. If we want to include the effect of the degenerate light
hole band the two expressions in Eqs.~(\ref{beff}) and (\ref{jeff})
should be multiplied by $2/3$ and $4/3$, respectively. By keeping the
lowest order in $J$ in Eqs.~(\ref{beff}) and (\ref{jeff}) we obtain
\begin{equation}
B_{eff}=\frac{|\Omega_{\sigma+}|^2-|\Omega_{\sigma-}|^2}{\delta^2}|\phi_{1s}|^2
v^2_0 J \frac{\hat{z}}{2}+ O(J^2)~,
\end{equation} and 
\begin{equation}
J_{eff}=
\frac{|\Omega_{\sigma+}|^2+|\Omega_{\sigma-}|^2}{\delta^2}|\phi_{1s}|^2
v^2_0 J^2 F_R/2 +O(J^3)
\end{equation} 
which recovers the Optical RKKY result of
Ref.~\onlinecite{piermarocchi02}. The magnetic field induced by
virtual excitons has recently been analyzed using a more fundamental
approach in the case of a single impurity by Combescot and
Betbeder-Matibet in Ref. \onlinecite{combescot04}. In this reference
the spin independent term that provides a correction to the optical
Stark shift is also discussed.

\section{Spin-spin coupling}
\label{app}
In this section we apply the results obtained above to (i) excitons
mediating the interaction between two electronic spins localized in
shallow donors (e.g. GaAs:Si) and (ii) excitons mediating the
interaction between two magnetic ions with spin 1/2 (two rare earth
ion Yb$^{3+}$ in InP).  Yb in InP is one of the most investigated rare
earth doped III-V material. In principle Yb$^{3+}$ in GaAs could be
used but it is technically more challenging to obtain samples where
only substitutional Yb is present.~\cite{taguchi90} We will focus on
the effect of the binding of excitons on the spin-spin coupling. The
parameters $J$ and the range of the potential $v_k$ can be fixed in
such a way that the single-spin exciton $T$-matrix reproduces the
binding energy and the spin configuration of the bound exciton
obtained from the experiment.

\subsection{Shallow donors}
For a scheme of the system we can refer again to Fig.~\ref{fig0}.  For
excitons interacting with a shallow neutral donor the effective mass
approximation can be used. The problem of excitons bound to neutral
donors and acceptors has been heavily investigated in the past
experimentally and theoretically.~\cite{dean79} In the case of GaAs it
is known that the exciton binds to a neutral donor with a binding
energy of about 1 meV.  It is also clear from the magnetic field
dependence of the bound exciton resonance that the two electrons are
paired in a singlet around the donor ion and the hole is bound by
Coulomb interaction. The picture is very similar to the one of a
positive charge bound to an H$^-$ ion. As in the H$^-$ ion case, the
dominant term responsible for the binding of the two electrons is a
kinetic exchange term and we can therefore disregard the effect of a
spin independent term in the Hamiltonian. The range of the kinetic
exchange is determined by the hybridization between the localized
electron state in the neutral donor and the electron state in the free
exciton. We therefore assume that the $v_k$ is of the form
\begin{equation}
v^2_k=\frac{1}{1+(\Lambda k)^2}~,
\end{equation}
where the parameter $\Lambda$ determines the range of the potential.
In the following we will use the excitonic atomic units where energy
is given in excitonic $Ry^*$ and lengths are in Bohr radius
a$_B^*$. $Ry^*=\frac{e^4 \mu}{ \varepsilon _0^2 2\hbar^2}$ where $\mu$
is the reduced mass of the electron hole system (in electronic mass
units) and $\varepsilon_0$ is the static dielectric constant in the
semiconductor. $a^*_B = \frac{\varepsilon_0 \hbar^2}{e^2 \mu}$, and the
relation $Ry^*=\frac{\hbar^2}{2 \mu {a^*_B}^2}$ holds.  For GaAs these
units give $1Ry^*\sim$ 5 meV and $a^*_B \sim $ 100 \AA. Using these
units we can calculate the functions $F_0$ and $F_R$ in Eq. (\ref{f0})
and Eq. (\ref{fr}) as
\begin{eqnarray}
  F_0&=&-\frac{1}{4 \pi \Lambda} \frac{1}{(\Lambda \sqrt{\delta
\nu}+\nu)}\\ 
F_R&=& -\frac{1}{4 \pi R} \frac{e^{-R/\Lambda}-e^{-R
\sqrt{\delta/\nu}}}{\Lambda^2 \delta -\nu}~,
\end{eqnarray}
where $\nu=\mu/M$ is the reduced total mass ratio of the excitonic
system which is about 1/5 in GaAs taking $m_e=0.08$ and $m_h=0.17$.
Notice that $F_R$ can be rewritten as 
\begin{equation}
F_R=F_0 ~\Lambda \nu
\frac{e^{-R/\Lambda}-e^{-R \sqrt{\delta/\nu}}}{R(\Lambda \sqrt{\delta
\nu} - \nu)} 
\end{equation}
and has no poles for positive detuning $\delta$; for $R \gg \lambda$, 
$F_R$ has a Yukawa form with a detuning-related decay length
$a_B^* \sqrt{\nu/\delta}$ as found in Ref.~\onlinecite{piermarocchi02},
while at short $R$ the finite range of the potential regularizes the
$1/R$ divergence.  In $v_k$ we take $\Lambda=$0.25. Using the fact
that the exciton binds to the donor only in the singlet channel, we
can determine the value of $J$ in the $T^S$ in Eq.~(\ref{ts}) in such a way
to have a pole at the experimental binding energy. The $J$ is positive, as
expected from the fact that the kinetic exchange is antiferromagnetic,
and we take its value to be $J=1 ~Ry^* (a^*_B)^3$ which gives a binding
energy for the singlet of 0.23 $Ry^*$, in accordance with the
experimental value of 1 meV. The triplet is unbound.

We plot in Fig.~\ref{fig1} the coupling constant $J_{eff}$ obtained
from Eq.~({\ref{jeff}) as a function of the energy of the laser
measured from the bottom of the free exciton band,
$\delta=\epsilon_0-\hbar\omega_L$. A small imaginary contribution to
the energy, $\eta=0.0001 Ry^*$, has been added in all the plots. The Rabi
energy is $\Omega_{\sigma+}=$0.05 meV. The $\sigma-$ component of the
Rabi energy is zero. The separation between the two neutral donors $R$
is 2 $a_B^*$. In the region of large detuning we have a ferromagnetic
coupling in agreement with the results obtained in the ORKKY
limit. When we approach the energy corresponding to the binding of the
exciton to the impurity, at $\delta=0.23$, we observe that the
interaction is strongly enhanced and there is a region with an
antiferromagnetic (AF) coupling. Multiple scattering between the two
impurities results in the formation of bonding-antibonding states for
the exciton. When the light has a frequency in the bonding-antibonding
gap the effective interaction changes sign. This is analogous to the
antiferromagnetic coupling generated by superexchange in magnetic
materials.~\cite{anderson63} When the laser is tuned above the
resonances we recover again the ferromagnetic coupling. In the same
plot we also show the effective coupling that would result by keeping
the lowest order in J (ORKKY). In this case no resonances due to the
binding of the excitons are present and, in order to obtain a sizeable
coupling, the laser has to be tuned close the bottom of the excitonic
band.
\begin{figure}
\centerline{\includegraphics[scale=0.9]{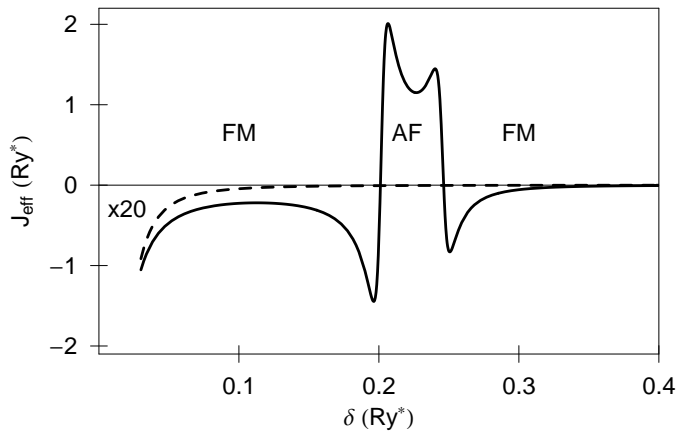}}
\caption{Coupling constant $J_{eff}$ between the two electronic spins
localized in a shallow neutral donor embedded in GaAs as a function of
the laser energy, measured from the bottom of the free exciton band.
The intensity of the laser corresponds to a Rabi energy
of $\Omega_{\sigma+}=0.05$ meV. The dashed line gives the result
predicted with the same parameters using second order perturbation
theory in the coupling constant $J$. \label{fig1}}
\end{figure}
In Fig.~\ref{fig2} we show a contour plot of the effective spin-spin
coupling as a function of the detuning and impurity separation.  Black
corresponds to strong negative and white to strong positive coupling,
the gray tone in the upper-right corner corresponds to zero. At large
$R$ the coupling is mostly ferromagnetic and there is only a small
region close to the exciton binding energy where the coupling can be
AF (white region in the plot). When the distance between the two
impurities decreases, the bonding-antibonding gap and the region
corresponding to the antiferromagnetic coupling is wider. The dashed
line indicates a change of sign of $J_{eff}$.  Notice also the
different decay of the interaction as a function of $R$ for different
values of the detuning. At $\delta=$0.4 the maximum strength is at
$R=0.8$ and decays quickly within a quarter of $a_B^*$ to the minimum
value in the plot. At $\delta=$0.1 the same minimum is reached within
a much larger interval of about 2 $a_B^*$. This is consistent with the
fact that at a small detuning there is a contribution from the free
exciton band which can give a longer range for the effective
interaction.
\begin{figure}
\centerline{\includegraphics[scale=0.85]{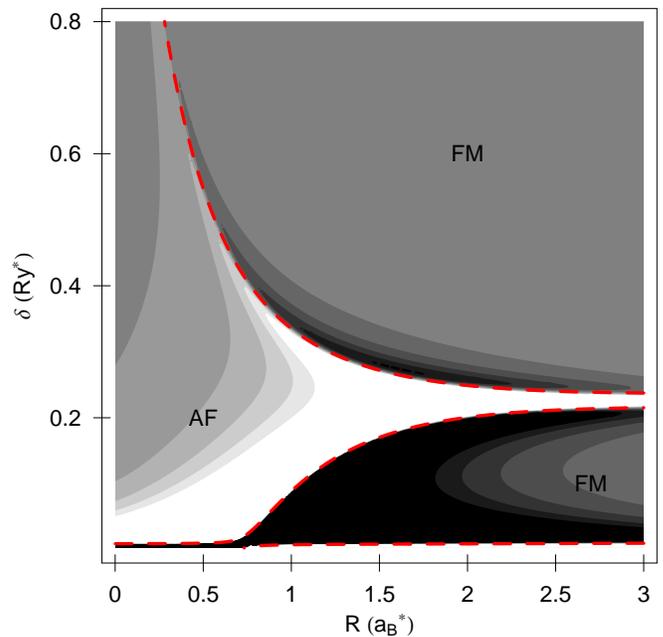}}
%\vspace{0.5 cm}
\caption{$J_{eff}$ as a function of the donors separation $R$ and
detuning $\delta $ . The contour plot identifies the regions where the
coupling is Ferromagnetic (FM) or antiferromagnetic (AF). The (red)
dashed lines indicate a change of sign of $J_{eff}$.\label{fig2}}
\end{figure}

\subsection{Rare earth impurities}
\label{Yb}
The magnetic properties of the Yb$^{3+}$ ion in
III-V~\cite{masterov90} arise from its partially filled 4$f$ shell,
possessing 13 electrons. In III-V materials, for a substitutional
impurity, the crystal fields split the ground manifold of the ion into
two doublets (spin=$\frac{1}{2}$), $\Gamma_6$ and $\Gamma_7$, and a
quadruplet (spin=$\frac{3}{2}$), $\Gamma_8$. The lowest lying state is
the Kramers doublet $\Gamma_6$, which behaves like a spin
$\frac{1}{2}$ with an effective isotropic $g=24/7$.~\cite{aszodi85} Yb
in InP replaces Indium and acts as an isoelectronic trap. From
electrical \cite{whitney88} and optical \cite{lambert90} measurements
it is known that the exciton binds to this isoelectronic impurity with
a binding energy of 30 meV.  The binding is due to a short range
potential that arises from the difference in the core pseudopotential
between the impurity and the host ion it
replaces.~\cite{baldereschi72} It is reasonable to assume that this
short range potential is spin-independent and we take it into account
by adding to the exciton-impurity Hamiltonian of Eq.~(\ref{h1}) the
term
\begin{equation}
H_2=\frac{1}{V}\sum_{k,k'}\Delta_{k,k'}(1+e^{-i(k'-k)R})b^{\dagger}_{k'
\alpha \beta} b_{k \alpha \beta}~.
\end{equation}
We use the separable potential approximation also for the
spin-independent short-range potential and we parametrize it in the
form $\Delta_{k,k'}=\Delta v_k v_{k'}$, i.e. it has the same $k$
dependence of $J_{k,k'}$. A more general analytical result can be
obtained using a separable form for $\Delta_{k,k'}$ with different
coefficients, but we expect the range of the $s$-$f$ exchange and that
of the impurity potential to be very similar.  The value of $\Delta$
is determined by imposing that the exciton-single impurity T-matrix
has a pole for both singlet and triplet channels at 30 meV. Following
the same procedure used in Sec.~\ref{one} we obtain for the T
operators in the singlet and triplet channels
\begin{subequations}
\label{STdelta}
\begin{eqnarray}
T^{S}&=&\frac{-3/4 J+\Delta}{1+ 3/4 J F_0-\Delta F_0} \\ T^{T}&=&
\frac{J/4+\Delta}{1- 1/4 J F_0-\Delta F_0}~.
\end{eqnarray}
\end{subequations}
The expressions for the $J_{eff}$ and $B_{eff}$ modified by the
presence of $\Delta$ can be obtained by plugging the
Eqs.~(\ref{STdelta}) in the general expressions of
Eqs.~(\ref{alphabeta}) in Appendix~\ref{appendix}. The quantity $J$ is
the $s$-$f$ exchange interaction between the impurity and the electron
in the exciton. In typical rare earth ferromagnetic semiconductors the
$s$-$f$ exchange is ferromagnetic and is of the order of few eV
\AA$^3$~~\cite{nolting80}, comparable to the $s$-$d$ exchange in Mn
based diluted magnetic semiconductors.~\cite{furdyna88} We are using
$J=-10^{-4}$ in our units which corresponds to a conservative estimate
of 0.7 eV \AA$^3$ in InP. In InP the value of the $Ry^*$ is about the
same as that of GaAs (5 meV), while the Bohr radius is about 120
\AA. For $\Lambda$ in $v_k$ we take $\Lambda=0.01$ which is of the
order of the ionic radius of Yb$^{3+}$.

We show in Fig.~\ref{fig3} the contour plot of $J_{eff}$ as a function
of the laser detuning $\delta=\epsilon_0-\hbar \omega_L$ and of the
separation between the impurities $R$. At large distances we observe
two resonances related to the binding of the exciton in the singlet
and triplet channels. Fig.~\ref{fig4}~(a) shows in detail the
$J_{eff}$ for a distance $R= 1 ~a_B^*$.  The two peaks in
Fig.~\ref{fig4}~(a) correspond to the exciton bound to the impurity in
the triplet and singlet channel.  The peak at larger detuning
corresponds to the triplet since the $s$-$f$ exchange is
ferromagnetic. For shorter distances we see from Fig.~\ref{fig3} that
each of the two peaks starts to split. The singlet (at smaller
detuning) follows a behavior similar to the one of the shallow donors
described above: the bonding and antibondig states identify a region
where the coupling become antiferromagnetic. The triplet state splits
in many different peaks as can be seen from Fig.~\ref{fig4}~(b). The
sign of the interaction can change many times as a function of the
detuning in this short distance region. This is indicated by the sign
of $J_{eff}$ plotted in the lower part of Fig.~\ref{fig4}~(b). Overall
the antiferromagnetic coupling dominates at short distances while the
interaction is ferromagnetic at large $R$.

\begin{figure}
\centerline{\includegraphics[scale=0.85]{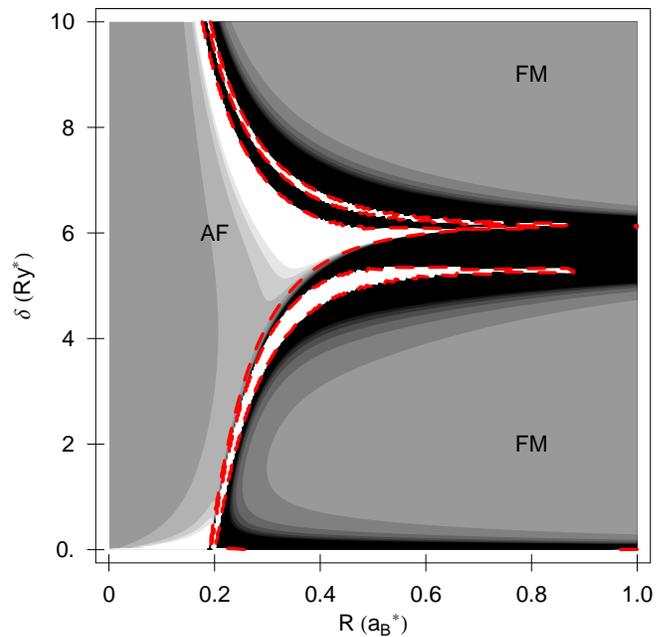}}
\caption{Coupling constant $J_{eff}$ between the two magnetic Yb$^{3+}$
localized in InP as a function of the laser detuning
$\delta=\epsilon-\hbar \omega_L$ and separation between the ions. The
dashed (red) lines indicate $J_{eff}$=0 and a change form
ferromagnetic to antiferromagnetic coupling.  The intensity of the
laser corresponds to a Rabi energy of $\Omega=0.1$ meV. \label{fig3}}
\end{figure}
 
\begin{figure}
\centerline{\includegraphics[scale=1.2]{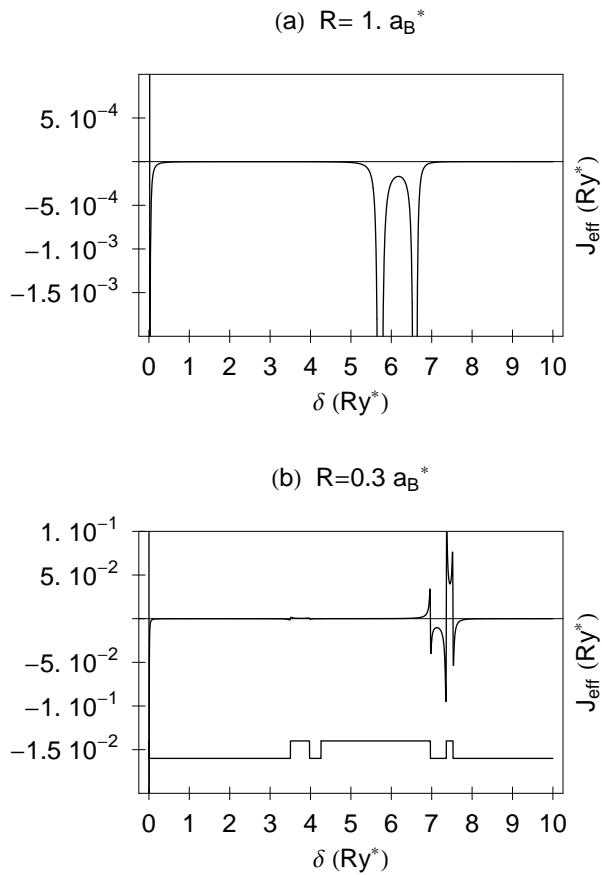}}
\vspace{0.5 cm}
\caption{Coupling constant $J_{eff}$ between two Yb$^{+3}$ ions in InP
as a function of the detuning $\delta$. (a) Large distance. The
coupling is ferromagnetic and the resonances in the interaction are
close to the energy of the exciton bound to the Yb. (b) Short
distance. The triplet channel splits in many different peaks
producing many changes of the spin-spin coupling sign. The lower curve
shows the sign of the coupling constant.  \label{fig4}}
\end{figure}

\section{Discussion}
\label{qc}
The spin-spin interaction control discussed in this paper has
potential applications in quantum computing implementations.  In fact,
an optical control of the spin of electrons localized in quantum dots
or impurities has several advantages with respect to approaches where
electrodes are needed. Ultrafast lasers are available, promising the
realization of quantum gates in time scales that are hard to achieve
with an electrical control. Laser are also very flexible from a
control perspective since pulse shaping can be used to advantage in
the accuracy and speed of quantum operations.~\cite{chen01} Finally,
metallic electrodes necessarily add a source of noise for the quantum
system, and they are not needed in an optical scheme. The possibility
of changing the sign of the spin-spin interaction exploiting the
exciton bound states can add flexibility to many control schemes for
the qubits, like e.g. in the only exchange scheme.~\cite{lidar00} We
have seen that resonances in the spin-spin coupling induced by the
binding of the excitons can increase the magnitude of the interaction
for distances that are reasonable from a nanofabrication point of
view. This will imply that lasers with lower intensities can be
employed in the control. The polarization of the light represents an
additional control parameter that can be used to selectively address
qubits with an optically induced magnetic field. This is also an
advantage from a practical point of view since it could simplify the
experimental setup by eliminating the need of an external magnetic
field.

Although the feasibility of single impurity spectroscopy in
semiconductors has been proven \cite{woggon01,strauf02}, little
attention has been paid to optical properties of impurity-bound
excitons for information storage and processing. Impurities deserve at
least the same attention as quantum dots for such applications. Their
homogeneous character and the variety of properties that one can
obtain combining different hosts and ions are indeed peculiar
advantages. An exciton bound to an impurity has optical properties
very similar to an exciton trapped in a shallow quantum dot.  Most of
the ideas involving excitons in quantum dots as a main ingredient for
quantum information and communication can be reformulated for excitons
bound to impurities.  We have provided  only two examples here,
but our phenomenological theory, being based on inputs from the
experiments, is very flexible and many other combinations of host and
ions can be used to explore a large range of confinement energy and
different optical properties. We also have seen that the spin-spin
coupling has a resonant behavior at frequencies depending on the
separation between the impurities. By organizing the impurities in
chains with different separation this can be used to selectively
address a single pair of impurities and it allows for scalability.

A very special case is represented by impurities in silicon. This
material has obvious technological advantages and many proposals for
using impurities in Si for quantum computing have been
suggested.~\cite{golding03,vrijen00,kane98} In particular, the optical
control of electronic spins localized by deep donors in Si using a
{\it control} impurity has been proposed.~\cite{stoneham03} In the
scheme we are suggesting here, the exciton bound to the impurity plays
the role of the  {\it control} impurity and it takes advantage of
the host material for mediating the interaction.  Even if Si is an
indirect gap material, there is a finite optical coupling to the
exciton bound to the impurity due to symmetry breaking. Additional
complications in the use of excitons bound to donors for mediating
spin-spin coupling arise form the valley degeneracy in
Si.~\cite{chang82} We will address the optical spin control of
impurities in Si and the role of valley degeneracy in a future
publication.

Excitons bound to rare earth magnetic ions can be controlled very
rapidly and efficiently due to their strong dipole moment. Their
dipole moment is mainly determined by the optical properties of the
host material, since it involves the creation of electron hole pairs
across the semiconductor gap. At the same time, they interact with the
internal degrees of freedom in the core $f$ states. Schemes involving
excitons bound to rare earth impurities in III-V materials bring in
the advantages of the optical properties of the host and the stability
of the internal degrees of freedom of the $f$ orbitals in the rare
earth ion where the qubit is stored. This hybrid system is thus
extremely powerful, providing both reliable storage and fast
processing of information.

Finally, the light controlled spin-spin coupling in a semiconductor
matrix is also appealing for the coherent control of macroscopic
properties of materials. This was the idea behind the coherently
induced ferromagnetism in Ref.~\onlinecite{fernandez03}. There, a
finite critical temperature for a paramagnetic to ferromagnetic
transition in diluted magnetic semiconductors was found when the
material is coupled to a strong laser field. The results presented in
this paper suggest that the presence of bound states could enhance the
effect.  Also, the same idea could be used in other systems
where the light can induce antiferromagnetic or glassy phases starting
form a paramagnetic system. This represents a unique opportunity to
study phase transitions in a solid where the coupling is controlled by
an external field and may open to a new class of {\it controlled}
materials to be investigated.

In conclusion, we have studied the problem of two spins $\frac{1}{2}$
localized by impurities in semiconductor in the presence of an intense
light field. The light induces a frequency dependent spin-spin
coupling and a magnetic field that can be controlled by the
polarization of the light. The effects are enhanced by the presence of
impurity bound excitons which may split in bonding antibonding states
in the case of two impurities. The sign of the spin-spin coupling is
generally ferromagnetic, but it can switch to antiferromagnetic when
the laser is tuned in the bonding-antibondig gap. We have developed a
flexible theoretical approach based on scattering theory where the
parameters from the experiment can be used to estimate the size of the
effect. We have discussed explicitly the case of two neutral donors in GaAs
and two rare earth magnetic ions (Yb$^{3+}$) in InP.

\acknowledgments This work was supported by NSF DMR-0312491. We thank
T. A. Kaplan for enlightening discussions on theory of magnetism, and
S. D. Mahanti for suggesting useful references on RKKY.

\appendix
\section{Matrix Representation of the $T$ operator}
\label{appendix}
Using the basis set $|s^A_z,s^B_z,s_z\rangle$ 
we obtain for the spin products $s^A \cdot s$, $s^B \cdot s$, $s^A\cdot s^B$ the matrices 
\begin{subequations}
\begin{eqnarray}
s^A \cdot s=
\begin{bmatrix}
\frac{1}{4}&0&0&0&0&0&0&0\\
0&-\frac{1}{4}&0&0&\frac{1}{2}&0&0&0\\
0&0&\frac{1}{4}&0&0&0&0&0\\
0&0&0&-\frac{1}{4}&0&0&\frac{1}{2}&0\\
0&\frac{1}{2}&0&0&-\frac{1}{4}&0&0&0 \\
0&0&0&0&0&\frac{1}{4}&0&0 \\
0&0&0&\frac{1}{2}&0&0&-\frac{1}{4}&0 \\
0&0&0&0&0&0&0&\frac{1}{4} 
\end{bmatrix}~~\\
s^B \cdot s=\begin{bmatrix}
\frac{1}{4}&0&0&0&0&0&0&0\\
0&-\frac{1}{4}&\frac{1}{2}&0&0&0&0&0\\
0&\frac{1}{2}&-\frac{1}{4}&0&0&0&0&0\\
0&0&0&\frac{1}{4}&0&0&0&0\\
0&0&0&0&\frac{1}{4}&0&0&0 \\
0&0&0&0&0&-\frac{1}{4}&\frac{1}{2}&0 \\
0&0&0&0&0&\frac{1}{2}&-\frac{1}{4}&0 \\
0&0&0&0&0&0&0&\frac{1}{4} 
\end{bmatrix}~~\\
s^A \cdot s^B=\begin{bmatrix}
\frac{1}{4}&0&0&0&0&0&0&0\\
0&\frac{1}{4}&0&0&0&0&0&0\\
0&0&-\frac{1}{4}&0&\frac{1}{2}&0&0&0\\
0&0&0&-\frac{1}{4}&0&\frac{1}{2}&0&0\\
0&0&\frac{1}{2}&0&-\frac{1}{4}&0&0&0 \\
0&0&0&\frac{1}{2}&0&-\frac{1}{4}&0&0 \\
0&0&0&0&0&0&\frac{1}{4}&0 \\
0&0&0&0&0&0&0&\frac{1}{4} 
\end{bmatrix}~.
\end{eqnarray}
\end{subequations}
By substituting these expressions in $\Upsilon^A$ and $\Upsilon^B$ and
then in Eq.~(\ref{finalT}), we obtain after matrix inversions and
multiplications an expression for $T=(1-F_R^2
\Upsilon^A \Upsilon^B)^{-1} \Upsilon^A
[1+F_R \Upsilon^B]+(A \rightleftharpoons B)$~. The traceless part of this matrix is
\begin{equation}
\label{TT}
\begin{bmatrix}
\frac{\beta+2\alpha}{4}&0&0&0&0&0&0&0\\
0&\frac{\beta-2\alpha}{4}&\frac{\alpha}{2}&0&\frac{\alpha}{2}&0&0&0\\
0&\frac{\alpha}{2}&-\frac{\beta}{4}&0&\frac{\beta}{2}&0&0&0\\
0&0&0&-\frac{\beta}{4}&0&\frac{\beta}{2}&\frac{\alpha}{2}&0\\
0&\frac{\alpha}{2}&\frac{\beta}{2}&0&-\frac{\beta}{4}&0&0&0 \\
0&0&0&\frac{\beta}{2}&0&-\frac{\beta}{4}&\frac{\alpha}{2}&0 \\
0&0&0&\frac{\alpha}{2}&0&\frac{\alpha}{2}&\frac{\beta-2\alpha}{4}&0 \\
0&0&0&0&0&0&0&\frac{\beta+2\alpha}{4}
\end{bmatrix}
\end{equation}
where $\alpha$ and $\beta$ can be conveniently expressed as a function
of the single impurity $T^T$ and $T^S$ operators in Eqs.~(\ref{ST}) or
Eqs.~(\ref{STdelta}) as 
\begin{subequations}
\label{alphabeta}
\begin{eqnarray}
\alpha&=&\frac{2(T^T-T^S) (T^T F_R+1)}{(T^T F_R -1)[F_R(T^S-T^T+2 F_R T^T T^S)-2]}~~~~~\\
\beta&=&-\alpha \frac{(T^T-T^S) F_R}{(F_R(T^T-T^S+2 F_R T^T T^S)-2)}~.
\end{eqnarray}
\end{subequations}
where we have dropped the \{A,B\} index since we are considering two
identical centers.  Notice that the matrix in
Eq.~(\ref{TT}) can be rewritten as
\begin{equation}
\alpha (s^A+s^B)\cdot s+ \beta s^A\cdot s^B~.
\end{equation}

\end{document}